\documentclass[aps,jcp,superscriptaddress,amsmath,amssymb]{revtex4-1}

\usepackage{natbib}
\setcitestyle{super}

\usepackage{graphicx}
\usepackage{dcolumn}
\usepackage{bm}

\newcolumntype{.}{D{.}{.}{8}}

\usepackage[utf8]{inputenc}
\usepackage{physics}
\usepackage{amsmath, amssymb}
\usepackage{tabularx,booktabs,array,dcolumn}
\usepackage{setspace}
\usepackage{vmargin}
\usepackage{xcolor}
\usepackage{graphicx,psfrag,subfigure}

\usepackage{float}
\usepackage[all]{xy}
\usepackage{multirow}

\usepackage[round-mode=places,round-precision=4]{siunitx} 

\usepackage[unicode]{hyperref}
\hypersetup{
   unicode=true,          
   plainpages=false,
   colorlinks=true,       
   citecolor=blue,        
}

\newcommand{\cm}{cm$^{-1}$}

\newcommand{\som}{SOM}

\def\tmL{\tilde{\mathcal{L}}}

\def\trans{\emph{trans}}
\def\cis{\emph{cis}}
\def\delocalized{\emph{delocalized}}

\newcommand{\bos}[1]{\pmb{#1}}

\def\kmax{k^\text{max}}

\usepackage{epstopdf}

\definecolor{cream}{RGB}{222,217,201}

\def\som{Supplementary Information}

\begin{document}

\title{%
Vibrational infrared and Raman spectrum of HCOOH from variational computations
}

\author{Gustavo Avila}
\affiliation{
ELTE, E\"otv\"os Lor\'and University, 
Institute of Chemistry, 
P\'azm\'any P\'eter s\'et\'any 1/A,
1117 Budapest, Hungary}

\author{Alberto Mart\'in Santa Dar\'ia}
\affiliation{
ELTE, E\"otv\"os Lor\'and University, 
Institute of Chemistry, 
P\'azm\'any P\'eter s\'et\'any 1/A,
1117 Budapest, Hungary}
\affiliation{Departamento de Química Física, University of Salamanca,
37008 Salamanca, Spain}

\author{Edit M\'atyus}
\email{edit.matyus@ttk.elte.hu}
\affiliation{
ELTE, E\"otv\"os Lor\'and University, 
Institute of Chemistry, 
P\'azm\'any P\'eter s\'et\'any 1/A,
1117 Budapest, Hungary}

\date{\today}
\begin{abstract}
  \noindent 
All vibrational energies of the (\trans-, \cis-, \delocalized-) formic acid molecule are converged up to 4500~\cm\ beyond the zero-point vibrational energy 
with the GENIUSH-Smolyak variational approach and using an \emph{ab initio} potential energy surface [D. P. Tew and W. Mizukami, J. Phys. Chem. A, 120,
9815–9828 (2016)].
Full-dimensional dipole and polarizability surfaces are fitted to points computed at the CCSD/aug-cc-pVTZ level of theory. Then, body-fixed vibrational dipole and polarizability transition moments are evaluated and used to simulate jet-cooled infrared and Raman spectra of HCOOH.
The benchmark-quality vibrational energy, transition moment, and wave function list will be used in further work in comparison with vibrational experiments, and in further rovibrational computations.
\end{abstract}

\maketitle 

\clearpage

\section{Introduction}
\noindent
This work is a continuation of Ref.~\citenum{DaAvMa22} and has been motivated by a systematic experimental effort \cite{MeSu19,nejad2020concerted,Ne22PhD,ChNe23} to measure the infrared and Raman vibrational spectrum of the formic acid monomer to set experimental benchmarks and to challenge computational quantum dynamics methodologies.\cite{TeMi16,RiCa18,AeCaRiBr20,NeSi21,KaMe22,KeLu22,AeScBr22,AeBrGa22} 

In the present work, we aim to make progress with a complete quantum dynamical 
characterization of this five-atomic molecule.
In the one hand, we elaborate on the basis pruning condition to be able to converge all vibrational energies up to and beyond the top of the \emph{cis-trans} isomerization barrier with a convergence error smaller than the uncertainty of the potential energy surface (PES) representation. As a result, the comparison of experiment and theory will allow assessment of the quality of the PES (and underlying approximations). In all computations presented in this work, we use the \emph{ab initio} PES developed by Tew and Mizukami \cite{TeMi16} and leave the comparison with the PES developed by Richter and Carbonni\`ere \cite{RiCa18} for future work (together with evaluation of the computations for the deuterated isotopologues).

Second, since our vibrational energy and wave function list includes all combination and overtone states converged to a level that their accuracy is limited only by the accuracy of the PES, it is necessary to predict which transitions may be visible in an experiment. For direct comparison with infrared and Raman spectra, we compute and fit electric dipole and dipole polarizability surfaces (over the coordinate range relevant for the vibrational dynamics).

We start with a summary of a physically motivated coordinate definition (Sec.~\ref{sec:coordef}) along the lines of Ref.~\citenum{DaAvMa22} (see also Ref.~\citenum{MaDaAv22}). 
Then, Sec.~\ref{sec:elpol} documents the computation and fitting of the dipole and polarizability surfaces, Sec.~\ref{sec:varvib} describes the improved vibrational computational details, and Sec.~\ref{sec:computed} presents the simulated infrared and Raman spectra for jet-cooled molecular beam experiments. All computed data and Fortran subroutines for the fitted property surfaces are deposited as \som.

\section{Physically motivated internal coordinates\label{sec:coordef}}
Valence coordinates, bond distances, bond angles, out-of-plane bending and torsional angles provide a good starting point for an efficient description of the quantum nuclear motion in molecules. 

\begin{figure}
  \begin{center}
    \includegraphics[scale=0.25]{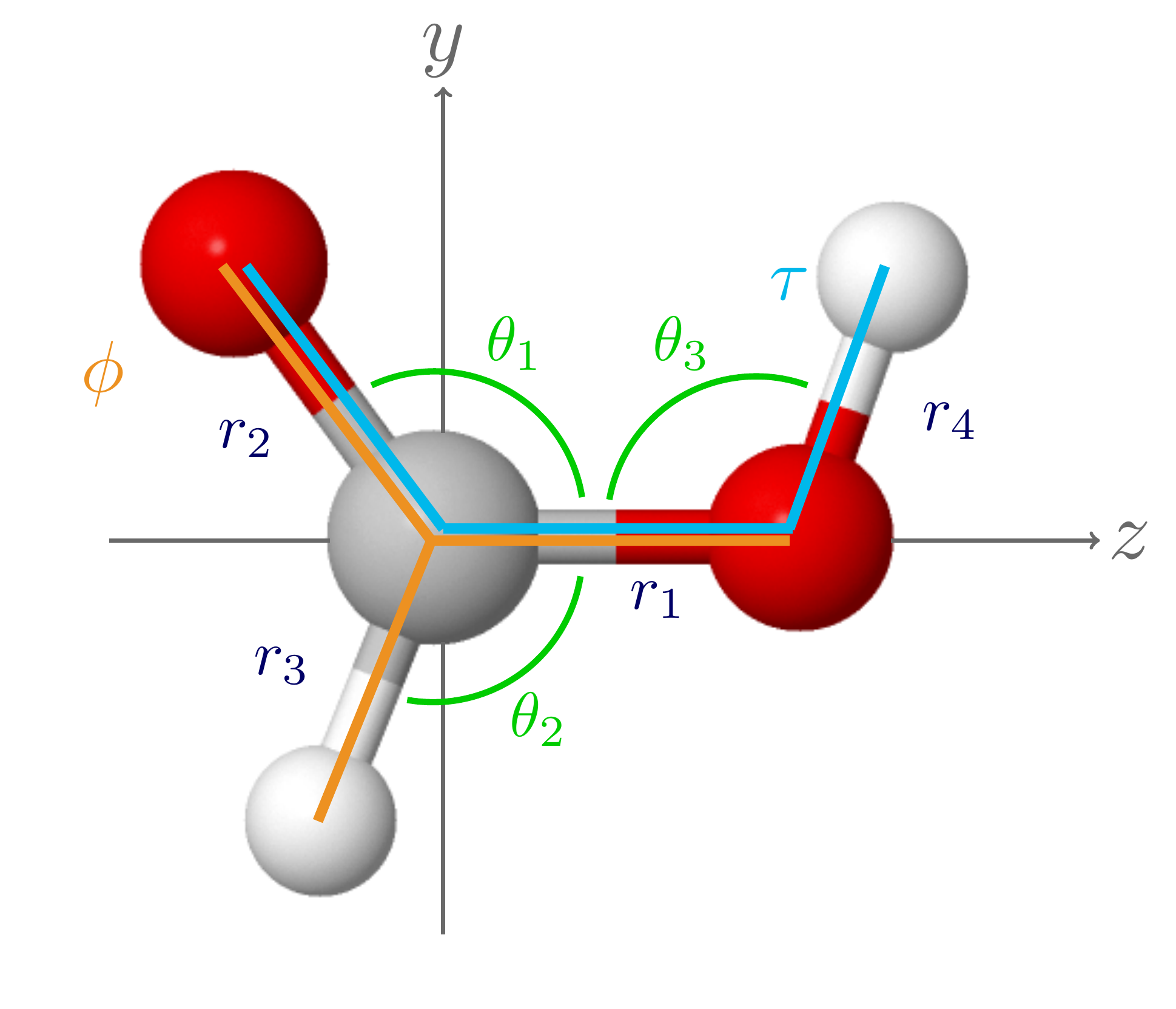}
    \end{center}      
    \caption{%
      Visualization of the `primitive' internal coordinates, Eq.~(\ref{eq:cartesian}), for the example of the \trans-formic acid molecule in its equilibrium structure. 
      }
    \label{fig:coord}
\end{figure}

Figure~\ref{fig:coord} summarizes the bond distance and angle coordinates used as `primitive' internal coordinates to describe vibrations of the formic acid.
Similarly to our earlier work,\cite{DaAvMa22} the $\xi_1,\ldots,\xi_9$  internal coordinates are defined by specifying the relation of  the body-fixed Cartesian coordinates and the internal coordinates as
\begin{align}
  &{\bos{r}}_{\text{C}_1}
  = 
  \bos{0} \; , 
\quad\quad\quad 
  \bos{r}_{\text{O}_2}
  = 
  \left(%
  \begin{array}{@{}c@{}}
    0 \\ 
    0 \\
    r_{1} \\
  \end{array}
  \right) \; ,
  \nonumber \\
  &\bos{r}_{\text{O}_1}
  =
  \left(%
  \begin{array}{@{}c@{}}
    0 \\
    r_{2} \cos{(\theta_{1}-\pi/2)} \\
   -r_{2} \sin{(\theta_{1}-\pi/2)}
  \end{array}
  \right) \; ,
  \nonumber \\
  &\bos{r}_{\text{H}_1}
  = 
  \left(%
  \begin{array}{@{}c@{}}
    r_{3} \cos{(\theta_{2}-\pi/2)} \sin\varphi\\
   -r_{3} \cos{(\theta_{2}-\pi/2)} \cos\varphi\\
   -r_{3} \sin{(\theta_{2}-\pi/2)}
  \end{array}\right) \; , 
  \nonumber \\
  &\bos{r}_{\text{H}_2}
  = 
  \bos{r}_{\text{O}_2} 
  + 
  \left(%
  \begin{array}{@{}c@{}}
    r_{4} \cos{(\theta_{3}-\pi/2)} \sin\tau \\
    r_{4} \cos{(\theta_{3}-\pi/2)} \cos\tau \\
    r_{4} \sin{(\theta_{3}-\pi/2)}
  \end{array}
  \right)
  \label{eq:cartesian}
\end{align}
with the $r_i\in[0,\infty)$ stretching, the $\theta_i\in[0,\pi]$ bending, the $\varphi\in[-\pi,\pi)$ out-of-plane bending, and the $\tau\in[0,2\pi)$ torsional coordinate domains.
So, the list of the `primitive' internal coordinates is
$(\xi_1,\ldots,\xi_9)=(r_1,r_2,r_3,r_4,\theta_1,\theta_2,\theta_3,\varphi,\tau)$
with the $\xi_9=\tau$ \cis-\trans\ torsional coordinate, the only large-amplitude degree of freedom in the system.

Similarly to our previous work,\cite{DaAvMa22} we define relaxed curvilinear normal  coordinates to describe the small-amplitude vibrations efficiently along the $\tau$ torsional path.
The small-amplitude coordinates are defined as
\begin{align}
  \xi_i(\tau)
  = 
  \xi^{\text{(eq)}}_i(\tau) 
  + 
  \sum_{j=1}^8
    \tmL_{i,j} (\tau)
    q_j  , 
    \quad i=1,\ldots,8 \; ,
  \label{eq:rcnormal}
\end{align}
where $q_j$ is the $j$th curvilinear normal coordinate. The $\xi^{\text{(eq)}}_i(\tau)$ `equilibrium' structure is obtained by minimization of the 8D cut of the PES for every $\tau$ values.
The $\tmL_{i,j} (\tau)$ linear combination coefficients are computed 
by the $\bos{GF}$ method over a grid of $\tau$ points. 
The grid of $\tau_k$ points included $k=1,\ldots,24$ values (equally spaced over the $[0,2\pi)$ interval) and 
the computed $\xi^{\text{(eq)}}_i(\tau_k)$ and $\tmL_{i,j} (\tau_k)$ values were interpolated using a Fourier expansion\cite{DaAvMa22} to obtain a functional form for the coordinate definition, Eq.~\eqref{eq:rcnormal}.
For the structure generation of the property surfaces and for most of the vibrational computations presented in this work, we use curvilinear normal coordinates with the $\tmL_{i,j}$ coefficients averaged for the $\tau=0^\circ$ (\trans) and $180^\circ$ (\cis) values similarly to the definition of Lauvergnat and Nauts,\cite{LaNa14}
\begin{align}
  \bos{\mathcal{L}}^\text{(ac)} 
  = 
  \frac{\bos{\mathcal{L}}(\tau=0^\circ) + \bos{\mathcal{L}}(\tau=180^\circ)}{2} \;,
  \label{eq:averL}
\end{align}
and
\begin{align}
  \bos{\xi}^{\text{(ac-eq)}}
  = 
  \frac{\bos{\xi}^{\text{(eq)}}(\tau=0^\circ) + \bos{\xi}^{\text{(eq)}}(\tau=180^\circ)  }{2} \; .
  \label{eq:averxi}
\end{align}
In the meanwhile, we also perform extensive test computations for the comparison of the efficiency of the averaged or relaxed ($\tau$-dependent) curvilinear normal coordinates (we note that there was a mistake in Ref.~\citenum{DaAvMa22} in relation with the (phase adjustment of) the relaxed, $\tau$-dependent coordinate definition). 

To calculate the $\xi_1,\ldots,\xi_8$ primitive internal coordinates from the $q_1,\ldots,q_8$ curvilinear internal coordinates, we use Eq.~\eqref{eq:rcnormal} and the mapping functions defined in Ref.~\citenum{DaAvMa22} to ensure that the resulting value of the (primitive) internal coordinate is in the mathematically correct interval.

\section{Electronic dipole and polarizability computations\label{sec:elpol}}
\subsection{Dipole and polarizability computational details\label{sec:dalton}}
To construct a representation for the electric dipole moment and dipole polarizability property surfaces,  a reasonable compromise was made in relation with the accuracy (level of electron correlation and size of the basis set) and the computational time. So, we decided to use the CCSD/aug-cc-pVTZ level of theory for the property computations over an extensive set of nuclear configurations \emph{(vide infra)}. 
The \emph{ab initio} computations have been carried out using the Dalton program package.\cite{Dalton} The polarizability matrix has been computed by using the Second-Order Polarization-Propagator Approximation (SOPPA) available in Dalton for the CCSD method.\cite{SOPPA} 
\begin{figure}
  \begin{center}
    \includegraphics[width=8cm]{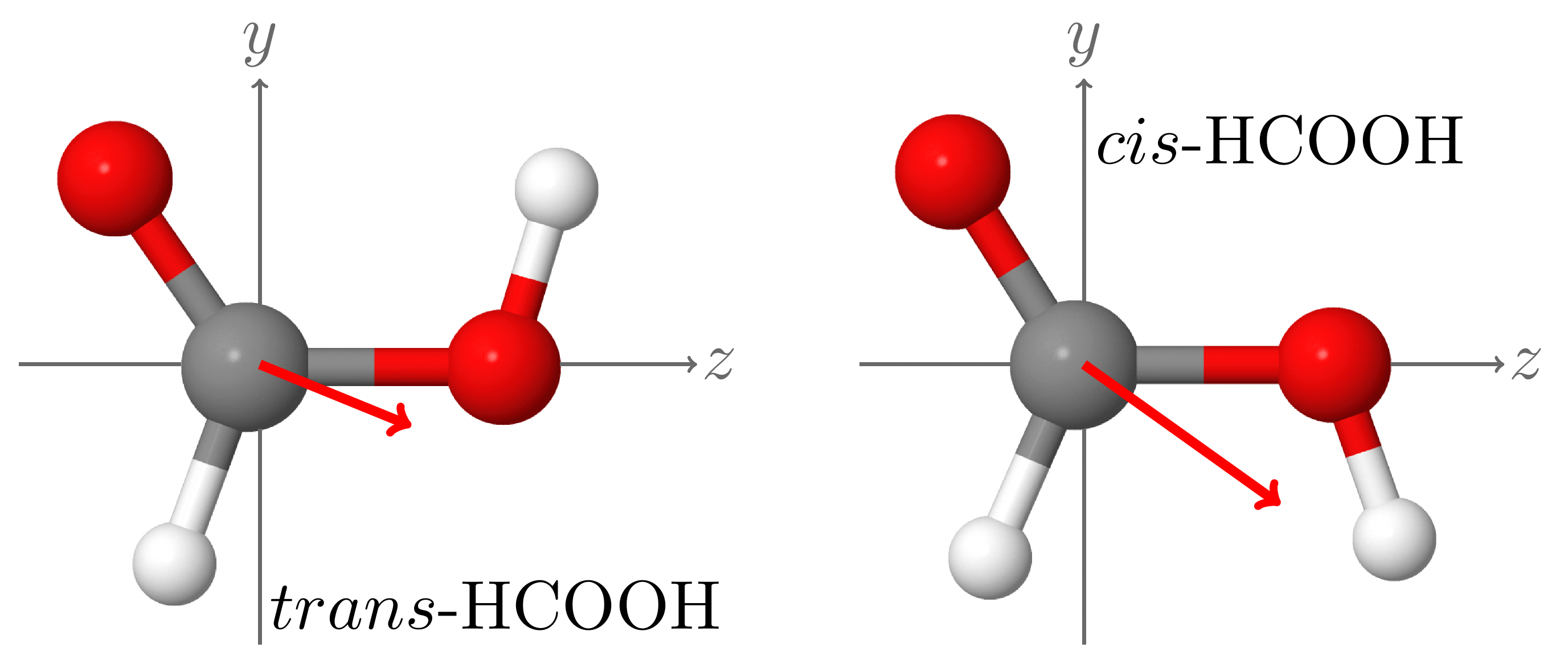}
  \end{center}    
  \caption{%
    Body-fixed frame and the formic acid molecule shown in both equilibrium configurations: $\trans$-HCOOH and $\cis$-HCOOH.
    The dipole moment vector is also shown (in red) in the figure for the two isomers. 
    \label{fig:structure}
    }
\end{figure}

\subsection{Dipole and polarizability surfaces}
The \emph{ab initio} values of the dipole moment vector and the dipole polarizability matrix elements computed by Dalton for a series of nuclear configurations have been fitted to a function of the (nine) internal coordinates of the system.

A simple fitting function (for components of the dipole vector or the polarizability matrix) could be defined as a direct-product of single-variable functions 
\begin{align}
  f^{\text{DP}}(\xi_1,\ldots,\xi_9) =
  \sum_{n_1=0}^{n_{\text{max}}} \cdots \sum_{n_9=0}^{n_{\text{max}}} 
    C_{n_1,\ldots,n_9}
    \prod_{k=1}^9
      F^{(k)}_{n_{k}}(\xi_k)  \; ,
  \label{eq:fit_dp}
\end{align}
where $f(\xi_1,\ldots,\xi_9)$ returns the value of the dipole moment ($f=\mu_i$ elements) or the dipole polarizability matrix ($f=\alpha_{ij}$ elements) for the $\xi_k\ (k=1,\ldots,9)$ values of the primitive internal coordinates.
To determine the linear combination coefficients of a direct-product-type representation, for instance with $n_\text{max}=7$, it would be necessary to compute at least $8^9\approx 134\times 10^6$ \emph{(ab initio)} data points, which is, of course, not feasible within a reasonable amount of time. 

Assuming that both the dipole and the polarizability surfaces are smooth functions of the (non-torsional) coordinates, we can reduce the number of points with respect to the (vibrational) dimensionality by fitting non-product functions without jeopardizing the accuracy of the representation. The strategy of truncating the direct product set of functions is similar to the construction of the non-product basis set in Ref.~\citenum{DaAvMa22}.
So, we define a non-product function as 
\begin{align}
  f^{\text{NON-DP}}&(\xi_1,\ldots,\xi_9) \nonumber \\
  &=
   \sum_{0\leq g(n_1,\ldots,n_8) \leq 7}
     C_{n_1,\ldots,n_8}
     \prod_{k=1}^8
     \left[%
       \xi_k-\xi_k^{(\text{eq})}
     \right]^{n_k}
     \sum_{n_9=0}^{n_9^\text{max}}
       F^{(9)}_{n_{9}}(\tau)  
  \label{eq:fit_nondp}
\end{align}
with the $g(n_1,\ldots,n_8) = n_1+n_2+n_3+n_4+n_5+n_6+n_7+n_8$ pruning function.
We note that pruning is applied only for the small-amplitude degrees of freedom, and the $\tau$ \cis-\trans\ torsional angle of HCOOH is not included in the restricted sum.
To represent the $\tau$ torsional dependence, we chose Fourier (sine and cosine) functions. 
To respect the plane symmetry of HCOOH ($C_s$ point group), sine or cosine functions are included depending on the polynomial order of the out-of-plane coordinate ($\xi_8$). 

Figure~\ref{fig:structure} shows the body-fixed frame definition used throughout this work. The OCO moiety is in the $yz$ plane and the C--O single bond defines the $z$ axis (Fig.~\ref{fig:structure}).

For the vibrational computations, we use 
pruned product functions of the curvilinear normal coordinates for the small-amplitude degrees of freedom and Fourier functions for the \cis-\trans\ torsional degree of freedom (Sec.~\ref{sec:vibmethod}).
To have an accurate representation of the dipole moment and dipole polarizability  over the dynamically relevant coordinate range, the coordinates of the vibrational computation are used to generate a grid of structures for the \emph{ab initio} computation and fitting. 
For a systematic generation of this grid of structures, we construct a set of pruned product grids.

To construct a pruned product grid, we first have to define a direct-product grid (the size of which grows exponentially with the vibrational degrees of freedom).
We label the collection of coordinate points for every vibrational degree of freedom by ${\mathcal{C}}$, so one point in the multi-dimensional grid can be labelled as
\begin{align}
  {\mathcal{C}}^{q_1}(k_1), 
  \ldots,  
  {\mathcal{C}}^{q_8}(k_8), 
  {\mathcal{C}}^{\tau}(k_\tau) \; .
\end{align}
Along the $\tau$ large-amplitude torsional degree of freedom, equally distributed points are defined and the plane symmetry with respect to 
$\tau=180^\circ$ is exploited. In this work, we used 
$\kmax_9=9$ points from the $[0,180]^\circ$ interval, \emph{i.e.,}
$\tau_{k_9}= (k_9-1) \cdot 180^\circ / (\kmax_9-1)=(k_9-1)\cdot 22.5^\circ$.
Regarding the small-amplitude degrees of freedom,
we used curvilinear normal coordinates (with coefficients averaged over the \cis\ and \trans\ parameterization) and the coordinate points were defined by the Gauss--Legendre (GL) quadrature points scaled to the $[-5.0,5.0]$ interval. Regarding the choice of the interval, it safely includes all points of the 15-point Gauss--Hermite quadrature typically used during the vibrational computations (for a more extended vibrational dynamics, additional points can be added to the property fit). Regarding the choice of the GL grid, it has a higher density of points near the edges of the interval, so a possibly good representation is provided near and (slightly) outside (extrapolation) the limits of the interval.

\begin{figure}
  \hspace{-0.35cm}\includegraphics[width=12cm]{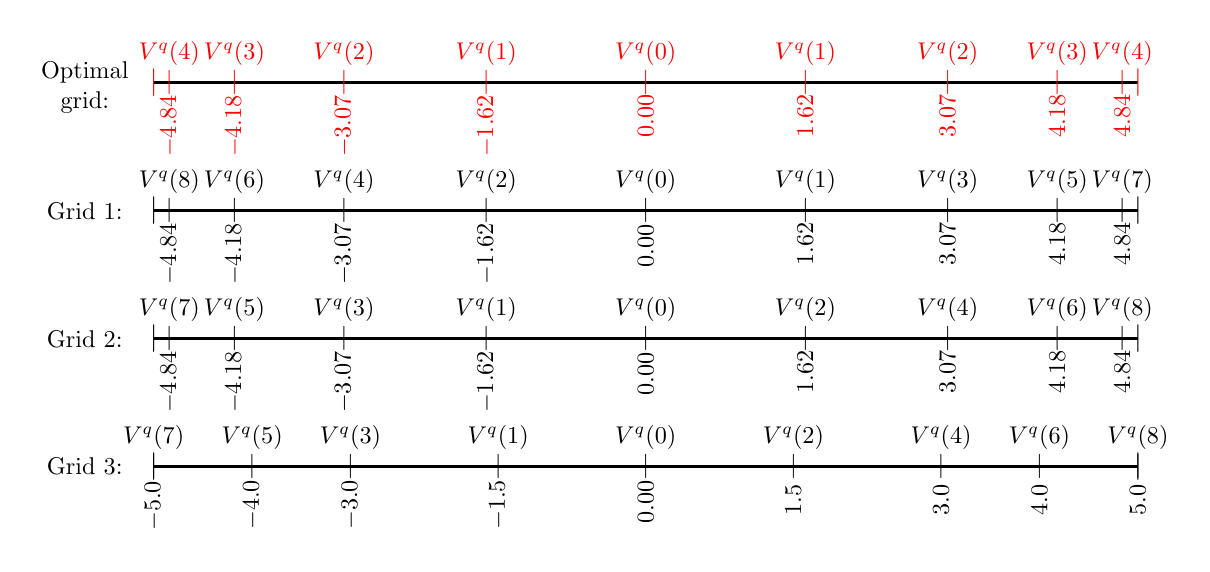}
  \caption{%
    Grid labelling types used for the construction of the multi-dimensional pruned coordinate grid for the \emph{ab initio} computations.
    \label{fig:gridpoints}
    }
\end{figure}

A simple direct-product 9-dimensional (9D) grid would be too large, 
so we want to cut this product grid by an appropriate pruning condition, which 
(significantly) reduces the number of points (for which the \emph{ab initio} computations are carried out), but provides a good coverage of the space relevant for the vibrational dynamics.
A simple truncation condition for the $q_i$ small-amplitude degrees of freedom is
\begin{align}
q_i (i=1,\ldots,8): \quad
& 0 \le k_1+k_2+k_3+k_4+k_5+k_6+k_7+k_8 \le k^{\text{max}}_q \nonumber \\ 
 \tau : \quad
& 0\le k_{\tau} \le k^{\text{max}}_\tau \; ,
\label{eq:fitprun}
\end{align}
while all points are retained for the $\tau$ torsional degree of freedom.
The resulting pruned grid depends on the labelling (enumeration) of the points in the one-dimensional grids. 
An ideal labeling, which ensures to the $\pm$ symmetry even in the pruned grid, is shown in the upper part of Fig.~\ref{fig:gridpoints} (`Optimal grid' with $k^\text{max}_q=4$ as an example). 
Unfortunately, a pruned grid with this labelling scheme would be too large, so a more economical construction is needed.

To proceed with the definition of a more compact multi-dimensional grid (for the small-amplitude vibrations), we first re-label the 1D grid points according `Grid 1' (Fig.~\ref{fig:gridpoints}). In the Grid~1 labelling, every index corresponds to only one point in the 1D grid. A 
multi-dimensional grid constructed from Grid~1-type 1D grids and 
the Eq.~\eqref{eq:fitprun} pruning would result in a (slight) over-representation of 
the positive-valued points.
Alternatively, we can define the `Grid~2' labelling (Fig.~\ref{fig:gridpoints}), which slightly favour the negative-valued points when used with the multi-dimensional  pruning condition in Eq.~\eqref{eq:fitprun}.

To have a multi-dimensional, but compact grid, we 
form the union of the Grid~1-type and the Grid~2-type multi-dimensional grids (G1 and G2), 
both generated separately with the pruning condition Eq.~\eqref{eq:fitprun} with
$k^\text{max}_q = 8$ and $k^\text{max}_\tau = 8$.
There are common points in the G1 and G2 grids, but every point is retained only once in the merged grid.

Finally, to improve the one-dimensional description, additional 1D grid points (with setting all other small-amplitude coordinates to the equilibrium value) are added for every small-amplitude degree of freedom to obtain the final multi-dimensional grid. The grid points in these one-dimensional grids are set manually (to approximately fill the `gaps' between the GL points in Grid~1 and 2), Grid~3 in Fig.~\ref{fig:gridpoints}.

As a result, a pruned multi-dimensional grid of $19\ 355\times 9=174\ 195$ distinct structures are generated and the \emph{ab initio} computations are performed for these structures. 

Although, for the structure generation, we used the $q_1,\ldots,q_8$ curvilinear normal coordinates (and $\tau$), the property  function, Eq.~\eqref{eq:fit_nondp},
is fitted using the $\xi_1,\xi_2,\ldots,\xi_8,\tau$ primitive internal coordinates. So, the resulting representation of the dipole and polarizability functions are mass independent and can be straightforwardly used (later) also for the isotopolgues.

The pruned fitting function, Eq.~\eqref{eq:fit_nondp}, contains $6\ 435\times9=57\ 915$ $C_{n_1,\ldots,n_8}$ linear combination coefficients (for every $\tau$ value) to be determined based on the generated \emph{ab initio} data. 
The functional representation of the properties is constructed in two steps. 
First, we fit the semi-rigid part, $C_{n_1,\ldots,n_8}$ for every value of $\tau$, 
which corresponds to solving an over-determined system of equations using the DGELS subroutine of LAPACK. 
Then, we fit these coefficients to a basis of cosine and sine functions,
\begin{align}
  C_{n_1,\ldots,n_8}(\tau) 
  = 
  \sum_{n=0}^8 A_n \cos(n\tau) \; ,
\label{eq:fitCcos}
\end{align}
and 
\begin{align}
  C_{n_1,\ldots,n_8}(\tau) 
  = 
  \sum_{n=0}^8 A_n \sin(n\tau)\; ,
  \label{eq:fitCsin}
\end{align}
depending on the symmetry of the $\mu_i$ and $\alpha_{ij}\ (i,j=x,y,z)$ elements and depending on the polynomial order (even or odd) of the out-of-plane bending coordinate.

In short, the algorithm used for generating the property functions is summarized as follows.
\begin{enumerate}
  \item 
    Select a multi-dimensional curvilinear normal coordinate point $(q_1,\ldots,q_8)$ and a $\tau$ point from the grid of structures.
  \item
    Calculate 
    the primitive internal coordinates $(\xi_1,\ldots,\xi_9)$ 
    at this $(q_1,\ldots,q_8,\tau)$ point   using the $\bos{\mathcal{L}}$ matrix and the $\xi_k^{(\text{eq})}$ reference structure, Eq.~\eqref{eq:rcnormal} (with the mapping functions defined in Ref.~\citenum{DaAvMa22}). 
  \item
    Calculate the Cartesian structure corresponding to $(\xi_1,\ldots,\xi_9)$
    in the body-fixed frame according to Eq.~\eqref{eq:cartesian} (Fig.~\ref{fig:structure}). 
  \item
    Compute 
    the $\bos{\mu}$ dipole moment vector and 
    the $\bos{\alpha}$ dipole polarizability matrix
    for this Cartesian structure
    using the Dalton program (with the computational details defined in Sec.~\ref{sec:dalton}).
  \item
    Repeat Steps 1--5 for every grid point. In our case, 19~355 grid points $(\xi_1,\ldots,\xi_8)$ for every $\tau_k$ values $(k=1,\ldots,9)$.
  \item
    Fit the 8D (small-amplitude) part of the property functions, Eq.~\eqref{eq:fit_nondp}, 
    using the \emph{ab initio} values computed in the previous steps, to obtain the $C_{n_1,\ldots,n_8}$ coefficients for every $\tau_k$ value (of the grid) using LAPACK.
  \item
    Repeat Step~6 for all $\tau_k$ grid points, $k=1,2,\ldots,9$ in the present work.
  \item
    Interpolate the $C_{n_1,\ldots,n_8}(\tau)$ coefficients for the $\tau_k$ points using Eqs.~\ref{eq:fitCcos} and \ref{eq:fitCsin} for every component of $\mu_i$ and $\alpha_{ij}$.
\end{enumerate}

We note that many $C_{n_1,\ldots,n_8}$ coefficients are zero by symmetry for $\tau=0^{\circ}$ and $\tau=180^{\circ}$.
If the $\mu_i$ or $\alpha_{ij}$ component is anti-symmetric (symmetric) with respect to the $\xi_8$ out-of-plane coordinate, then we have 4~352 (2~083) non-vanishing coefficients.

The absolute value of the difference of the fitted function and the \emph{ab initio} value corresponding to the grid points is used as a measure of the quality of the fit ($\eta$). Considering every $\tau$ value and every property component, the largest deviation is $\eta=3\times10^{-4}$ (in atomic units).
To assess the quality of the final property functions obtained by interpolating the $C_{n_1,\ldots,n_8}(\tau_k)$ coefficients over the $\tau_k$ grid, 
we have performed additional 100 \emph{ab initio} computations carried out at randomly selected structures. With respect to this test set, the largest absolute deviation (among all component functions) is $\eta=5\times 10^{-3}$, which is larger than the fitting error of the 8-dimensional cuts. 
Thus, the quality of the final property surfaces could be improved by adding more torsional points and sine/cosine functions in the final interpolation, but the present setup is already convenient for our purposes. (In the \som, we deposit all computed \emph{ab initio} points and structures for possible further use.)

Next, it is appropriate to introduce quantities which are invariant with respect to the selection of the body-fixed frame. 
For the dipole moment, we define its length,
\begin{align}
    |\mu| = \sqrt{\mu_x^2 + \mu_y^2 + \mu_z^2} \; .
\label{eq:dipole}
\end{align}
For the $3\times 3$ polarizability  matrix, the so-called isotropic and anisotropic polarizability are usually defined\cite{Lo77,NeReKiHe02} as 
\begin{align}
    a = \frac{1}{3} (\alpha_{xx} + \alpha_{yy} + \alpha_{zz})
\label{eq:pol_para}
\end{align}
and 
\begin{align}
    \gamma^2 = 
    &\frac{1}{2} \big[ 
    (\alpha_{xx}-\alpha_{yy})^2 + (\alpha_{xx}-\alpha_{zz})^2 + (\alpha_{yy}-\alpha_{zz})^2 \big] \nonumber \\
    & + 3\big[
    \alpha_{xy}^2 + \alpha_{xz}^2 + \alpha_{yz}^2
    \big] \; ,
\label{eq:pol_perp}
\end{align}
respectively.

\begin{figure}
  \begin{center}
    \includegraphics[width=8cm]{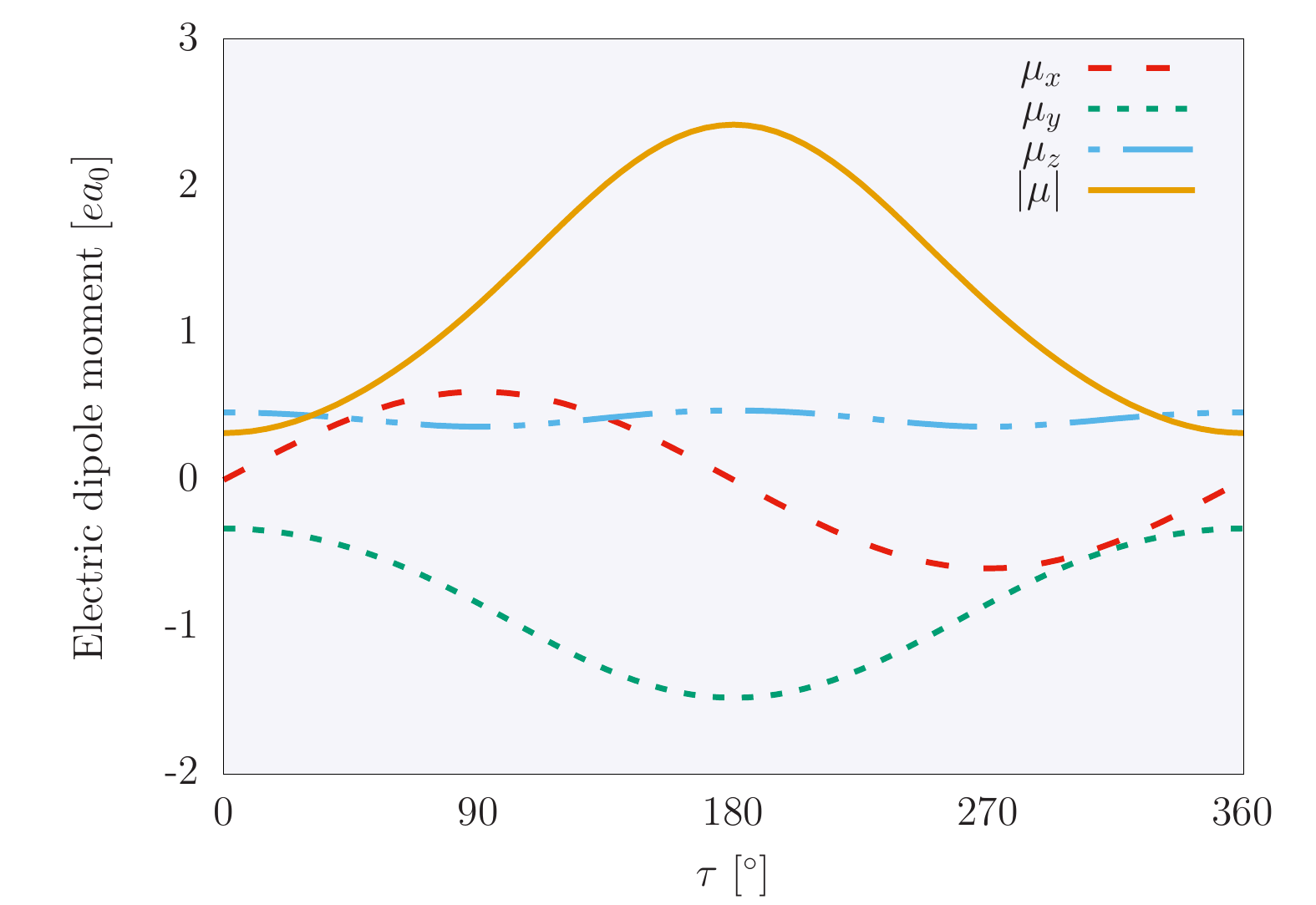}
  \end{center}    
  \caption{%
    Length of the electric dipole moment vector, $|\mu|$, as a function of the $\tau$ torsional angle (all other internal coordinates are fixed at their averaged \cis-\trans\  equilibrium value). The $\mu_x$, $\mu_y$, and $\mu_z$ components are also shown corresponding to the body-fixed frame defined in Fig.~\ref{fig:coord} and Eq.~\eqref{eq:cartesian}. 
    \label{fig:mu_tau}
    }
\end{figure}

The $|\mu|$ dipole length (and also the $\mu_x,\mu_y,\mu_z$ Cartesian components in our body-fixed frame) are shown along the torsional $\tau$ coordinate  in Fig.~\ref{fig:mu_tau} (while all other internal coordinates are fixed at the \cis-\trans\ averaged equilibrium value).
Since the equilibrium structure of the molecule is planar, and in our body-fixed frame definition (Fig.~\ref{fig:structure}), it is placed in the $yz$-plane corresponding to $\tau=0^{\circ}$ and $\tau=180^{\circ}$, the $\mu_y$ and $\mu_z$ component, as well as, the $|\mu|$ length is symmetric, whereas $\mu_x$ is antisymmetric, 
with respect to the $\tau=0^\circ$ and $\tau=180^\circ$ values. 

\begin{figure}
  \begin{center}
    \includegraphics[width=8cm]{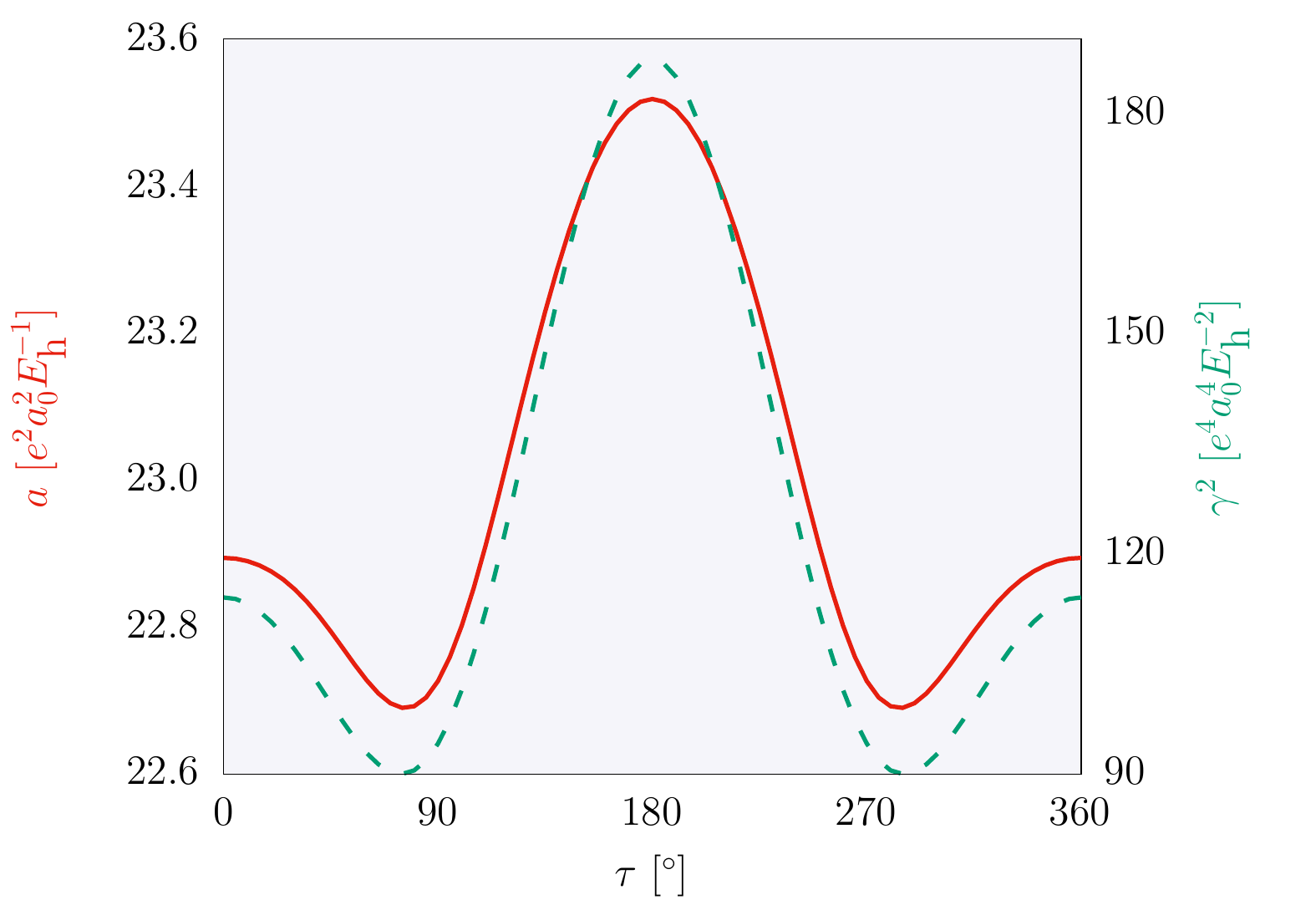}
  \end{center}    
  \caption{%
    Isotropic and anisotropic polarizabilities, $a$ (red, solid line) and $\gamma^2$ (green, dashed line), with respect to the $\tau$ torsional angle (all other internal coordinates are fixed at their averaged \cis-\trans\ equilibrium value).
    \label{fig:al_tau} 
    }
\end{figure}

Figure~\ref{fig:al_tau} shows the isotropic and anistropic parts of the polarizability, Eqs.~\ref{eq:pol_para} and \ref{eq:pol_perp}, with respect to variation of the $\tau$ torsional angle 
(while all other internal coordinates are fixed at the \cis-\trans\ averaged equilibrium value).

The coordinate-system invariant quantities are used to compute the vibrational transition moments and to simulate the jet-cooled infrared and Raman spectra in Sec.~\ref{sec:computed}.
The full dipole vector and the polarizability matrix will be used with the rovibrational wave functions to compute infrared and Raman rovibrational transition moments\cite{DaAvMa21a} for comparison with high-resolution spectroscopy experiments in future work.

\section{Variational vibrational computations\label{sec:varvib}}
\subsection{Vibrational methodological details\label{sec:vibmethod}}
The vibrational Hamiltonian corresponding to the curvilinear normal coordinates and the $\tau$ coordinate as defined in Sec.~\ref{sec:coordef} was considered in its `fully rearranged' form,\cite{AvMa19,AvMa19b,AvPaCzMa20}
\begin{align}
  \hat{H}^\text{v} 
  = 
  -\frac{1}{2} 
  \sum_{k=1}^{D} \sum_{l=1}^{D} 
    G_{kl} \frac{\partial}{\partial \xi_{k}}\frac{\partial}{\partial \xi_{l}}
  -\frac{1}{2} 
  \sum_{l=1}^{D} B_{l}
    \frac{\partial}{\partial \xi_{l}}
  +U
  +V \; ,
  \label{eq:hamiltonian}
\end{align}
where
\begin{align}
  B_{l} 
  = 
  \sum_{k=1}^{D} 
  \frac{\partial}{\partial \xi_{k}} G_{kl}  \; ,
\end{align}
and the pseudo-potential term is
\begin{align}
  U= \frac{1}{32} \sum_{k=1}^{D} \sum_{l=1}^{D}
    \Bigg[
    \frac{G_{kl}}{\tilde{g}^2} \frac{\partial\tilde{g}}{\partial \xi_k}
    \frac{\partial\tilde{g}}{\partial \xi_l}
    +4\frac{\partial}{\partial \xi_k} 
    \Bigg(
    \frac{G_{kl}}{\tilde{g}}
    \frac{\partial\tilde{g}}{\partial \xi_l}
    \Bigg)
    \Bigg] \;  .
  \label{eq:pseudopot}
\end{align}
To compute the vibrational eigenstates of this Hamiltonian, we constructed 
its finite basis representation.
The terms $G_{kl}$, $B_l$, and $U$ depend only on the vibrational coordinates 
and were computed by the numerical kinetic energy operator approach in the GENIUSH program\cite{MaCzCs09} over the integration grid \emph{(vide infra)}.

\begin{figure}
    \centering
    \includegraphics[scale=0.45]{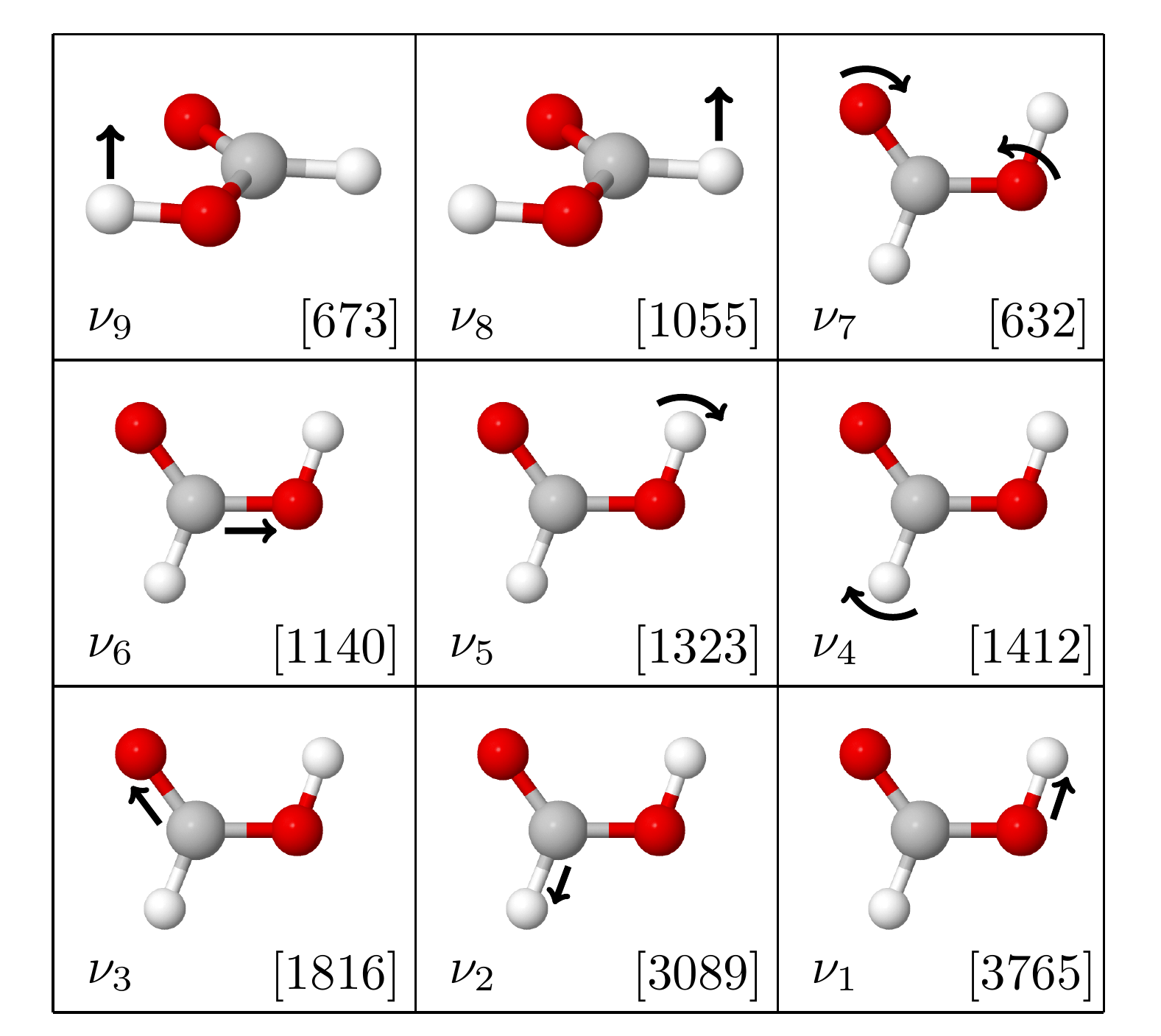}
    \caption{%
      Visualization of the normal modes corresponding to the global minimum (\trans) of HCOOH structure. The harmonic frequencies, in \cm, are shown in brackets. 
      Reproduced from A. Mart\'in Santa Dar\'ia, G. Avila, and E. Mátyus, J. Mol. Spectrosc. 385, 111617 (2022)
      (published under the Creative Commons CC-BY-NC-ND license).
    \label{fig:nctrans}
    }
\end{figure}

Regarding the basis functions, the dimensionless curvilinear normal coordinates were described with harmonic oscillator basis functions,
and Fourier (sine and cosine) functions were used for the $\tau$ coordinate.
Figure~\ref{fig:nctrans} shows the enumeration of the normal modes (and the corresponding harmonic frequencies) corresponding to \trans-HCOOH. The same ordering is followed also for the curvilinear normal coordinates.

Regarding the torsional functions for $\tau$, the Fourier basis was used to solve the 1-dimensional (1D) torsional Schrödinger equation,
\begin{equation}
  \label{eq:1dhamil}
  \hat{H} 
  = 
  -G_{\tau\tau} \frac{\partial^2}{\partial \tau^2} 
  - \frac{\partial G_{\tau\tau}}{\partial \tau} \frac{\partial}{\partial \tau} 
  + V_\tau \; .
\end{equation}
The computed torsional eigenfunctions can be labelled as \cis, \trans\ and \delocalized\ (Ref.~\citenum{DaAvMa22}) and were used to build the 9-dimensional (9D) basis set. 

To make the 9D variational computations feasible for a large number of vibrational states, we used basis and grid pruning techniques in Ref.~\citenum{DaAvMa22}, relying on the Smolyak scheme\cite{AvCa09,AvCa11,AvCa11b}
and its implementation in the GENIUSH program.\cite{AvMa19,AvMa19b}
In Ref.~\citenum{DaAvMa22}, we used the simplest basis pruning condition to reduce the basis set for the small-amplitude vibrations, 
\begin{align}
  P^\text{b}_0:\quad
  0 \le n_1+n_2+n_3+n_4+n_5+n_6+n_7+n_8 \le b \; ,
  \label{eq:old_basispruning}
\end{align}
which, with the $b=9$ pruning parameter, was sufficient to
converge (most) vibrational energies within 2~\cm\ up to 3500~\cm\ beyond  the zero-point vibrational energy (ZPVE), with some exceptions, which primarily corresponded to overtones of the $v_7$ OCO bending mode.

For simplicity, we use the curvilinear normal coordinates corresponding to the averaged coefficients, Eqs.~\eqref{eq:averL}--\eqref{eq:averxi},\cite{LaNa14} in this work.
We note that the plane symmetry is strictly retained for the averaged curvilinear normal coordinate representation, whereas it is perfectly recovered only for the converged results if relaxed curvilinear normal coordinates are used.
We have also studied the effect of using relaxed curvilinear normal coordinates, but so far, the advantage of relaxation appears to be small (for the present system and energy range).

In Ref.~\citenum{DaAvMa22}, we observed that (high) overtones and their combination bands of the $\xi_7$ O--C--O bending mode converges slowly. 
This vibrational degree of freedom appears to be the most anharmonic among the eight non-torsional degrees of freedom. To improve the convergence of the vibrational states, it is necessary to increase the number of basis functions for this degree of freedom.
Based on these observations, improved pruning conditions are considered in this work, which can be generally formulated as
\begin{align}
    0 \le \sum_{k=1}^8 G^k (n_k) \le b \; .
\label{eq:new_basispruning}
\end{align}
We use $G^k (n_k)=n_k$ identity for $k=1,2,\ldots,6,8$ (as in Ref.~\cite{DaAvMa22}),
but for $k=7$, we define a more elaborate mapping (which we call $P_1$ pruning, for short),
\begin{align}
P^\text{b}_1:\quad
&G^{7}(0)=0,\;
G^{7}(1)=1,\;
G^{7}(2)=2,\;
G^{7}(3)=3, \;
G^{7}(4)=3,\nonumber \\
&G^{7}(5)=5, \;
G^{7}(6)=5, \;
G^{7}(7)=7, \;
G^{7}(8)=7, \;
G^{7}(9)=8,\nonumber \\
&G^{7}(10)=8, \; 
G^{7}(11)=9, \;
G^{7}(12)=9, \;
G^{7}(13)=10, \nonumber \\
&G^{7}(14)=11, \;
G^{7}(15)=12, \;
G^{7}(16)=13 \;,
\label{eq:basis7}
\end{align}
which allows more basis functions for that degree of freedom.
Even more functions are added to the $k=7$ degree of freedom by defining (which we call $P_2$ pruning, for short)
\begin{align}
P^\text{b}_2:\quad
&G^{7}(0)=0,\;
G^{7}(1)=1,\;
G^{7}(2)=1,\;
G^{7}(3)=2, \;
G^{7}(4)=2,\nonumber \\
&G^{7}(5)=3, \;
G^{7}(6)=3, \;
G^{7}(7)=4, \;
G^{7}(8)=4, \;
G^{7}(9)=5,\nonumber \\
&G^{7}(10)=5, \; 
G^{7}(11)=6, \;
G^{7}(12)=6, \;
G^{7}(13)=7, \nonumber \\
&G^{7}(14)=7, \;
G^{7}(15)=8, \;
G^{7}(16)=8, \;
G^{7}(17)=9, \; 
G^{7}(18)=9, \nonumber \\
&G^{7}(19)=10, \;
G^{7}(20)=10, \;
G^{7}(21)=11, \;
G^{7}(22)=11 . \;
\label{eq:basis8}
\end{align}

\begin{table}
    \caption{%
      Number of multi-dimensional basis functions and grid points for three pruning conditions. 
      \label{tab:prun}
    }
    \centering
    \begin{tabular}{@{}r@{\ \ }c r@{\ \ } r@{\ \ \ } r @{}}
      \hline\hline\\[-0.35cm]
      \multicolumn{2}{c}{Pruning parameters} & & \multicolumn{2}{c}{Number of} \\
      \cline{1-2}\cline{4-5}\\[-0.35cm]
        $b$    &    $H$  & & \multicolumn{1}{c}{basis functions} & \multicolumn{1}{c}{grid points} \\
      \hline\\[-0.25cm]    
    \multicolumn{5}{l}{$P_0$ pruning condition, Eqs.~\ref{eq:old_basispruning} and \ref{eq:old_gridpruning}, used in Ref.~\citenum{DaAvMa22}:}\\
         8 & 19 & & 707 850   & 42 223 623 \\
         9 & 20 & & 1 337 050 & 72 656 063 \\
         10& 21 & & 2 406 690 & 132 043 839\\
         \hline\\[-0.25cm]  
    \multicolumn{5}{l}{$P_1$ pruning condition, Eqs.~\ref{eq:new_basispruning}--\ref{eq:basis7} and \ref{eq:new_gridpruning}--\ref{eq:new_gridmapping1} (this work):} \\
         10 & 21 & &  2 535 500    & 211 829 257 \\
         11 & 22 & &  4 399 725    & 374 873 881 \\
      \hline\\[-0.25cm]             
    \multicolumn{5}{l}{$P_2$  pruning condition, Eqs.~\ref{eq:new_basispruning}--\ref{eq:basis8} and \ref{eq:new_gridpruning}--\ref{eq:new_gridmapping2} (this work):} \\
         10 & 21 & & 3 743 740 &  213 924 337   \\
         11 & 22 & & 6 563 700 &   379 171 797    \\         
      \hline\hline
    \end{tabular}
\end{table}

A multi-dimensional quadrature grid is used to evaluate the multi-dimensional integrals, and it is also pruned in order to attenuate the computational costs. The grid pruning condition must be adapted to the basis pruning condition to ensure accurate potential and kinetic energy integrals for all retained multi-dimensional basis functions.
In our previous work,\cite{DaAvMa22} the pruning function for the Smolyak multi-dimensional grid was 
\begin{align}
P^\text{g}_0:\quad 8 \le i_1+i_2+i_3+i_4+i_5+i_6+i_7+i_8 \le H \;.
\label{eq:old_gridpruning}
\end{align}
In this work, a more elaborate `mapping' function is defined for the grid pruning, similarly to the improved basis pruning, Eq.~\ref{eq:new_basispruning}, as
\begin{align}
    8 \le \sum_{k=1}^8 g^k (n_k) \le H 
\label{eq:new_gridpruning}
\end{align}
where $g^k (n_k) = n_k\ (k=1,2,\ldots,6,8)$ (as in Ref.~\cite{DaAvMa22}), but for $k=7$, the following grid mapping functions are defined, corresponding to the Eqs.~\eqref{eq:basis7} and \eqref{eq:basis8} basis pruning functions, 
\begin{align}
P^\text{g}_1:\quad 
&g^{7}(1)=1, \;
g^{7}(2)=2, \;
g^{7}(3)=2, \;
g^{7}(4)=3, \;
g^{7}(5)=3, \nonumber \\
&g^{7}(6)=4, \;
g^{7}(7)=4, \;
g^{7}(8)=5, \;
g^{7}(9)=5, \;
g^{7}(10)=7, \nonumber \\
&g^{7}(11)=7, \;
g^{7}(12)=8, \; 
g^{7}(13)=8, \;
g^{7}(14)=8, \; \nonumber \\
&g^{7}(15)=9, \;
g^{7}(16)=9, \;
g^{7}(17)=9, \;
g^{7}(18)=10, \nonumber \\
&g^{7}(19)=10, \;
g^{7}(20)=12, \;
g^{7}(21)=12, \nonumber \\
&g^{7}(22)=14, \;
g^{7}(23)=14 \; .
\label{eq:new_gridmapping1}
\end{align}
and
\begin{align}
P^\text{g}_2:\quad &g^{7}(1)=1, \;
g^{7}(2)=2, \;
g^{7}(3)=2, \;
g^{7}(4)=3, \;
g^{7}(5)=3, \nonumber \\
&g^{7}(6)=4, \;
g^{7}(7)=4, \;
g^{7}(8)=5, \;
g^{7}(9)=5, \;
g^{7}(10)=6, \nonumber \\
&g^{7}(11)=6, \;
g^{7}(12)=7, \; 
g^{7}(13)=7, \;
g^{7}(14)=8, \; \nonumber \\
&g^{7}(15)=8, \;
g^{7}(16)=9, \;
g^{7}(17)=9, \;
g^{7}(18)=10, \nonumber \\
&g^{7}(19)=10, \;
g^{7}(20)=10, \;
g^{7}(21)=11, \nonumber \\
&g^{7}(22)=11, \;
g^{7}(23)=12 \; ,
\label{eq:new_gridmapping2}
\end{align} 
respectively.

As a general rule, 
the grid pruning function was designed to allow exact integration of a polynomial of maximal degree of five even for the highest `excited' (order) basis functions in the pruned basis set. For lower-excited basis functions a higher maximal polynomial degree (formally representing a Hamiltonian term) is integrated exactly.
All in all, these parameters ensure highly accurate integration of the Hamiltonian matrix elements, resulting in a practically variational approach (the energies converge to the exact value from `above').

For the torsional coordinate, we use the 55 basis functions and 79 trapezoidal quadrature points as defined in our previous work.\cite{DaAvMa22}

\begin{figure}
  \includegraphics[width=11cm]{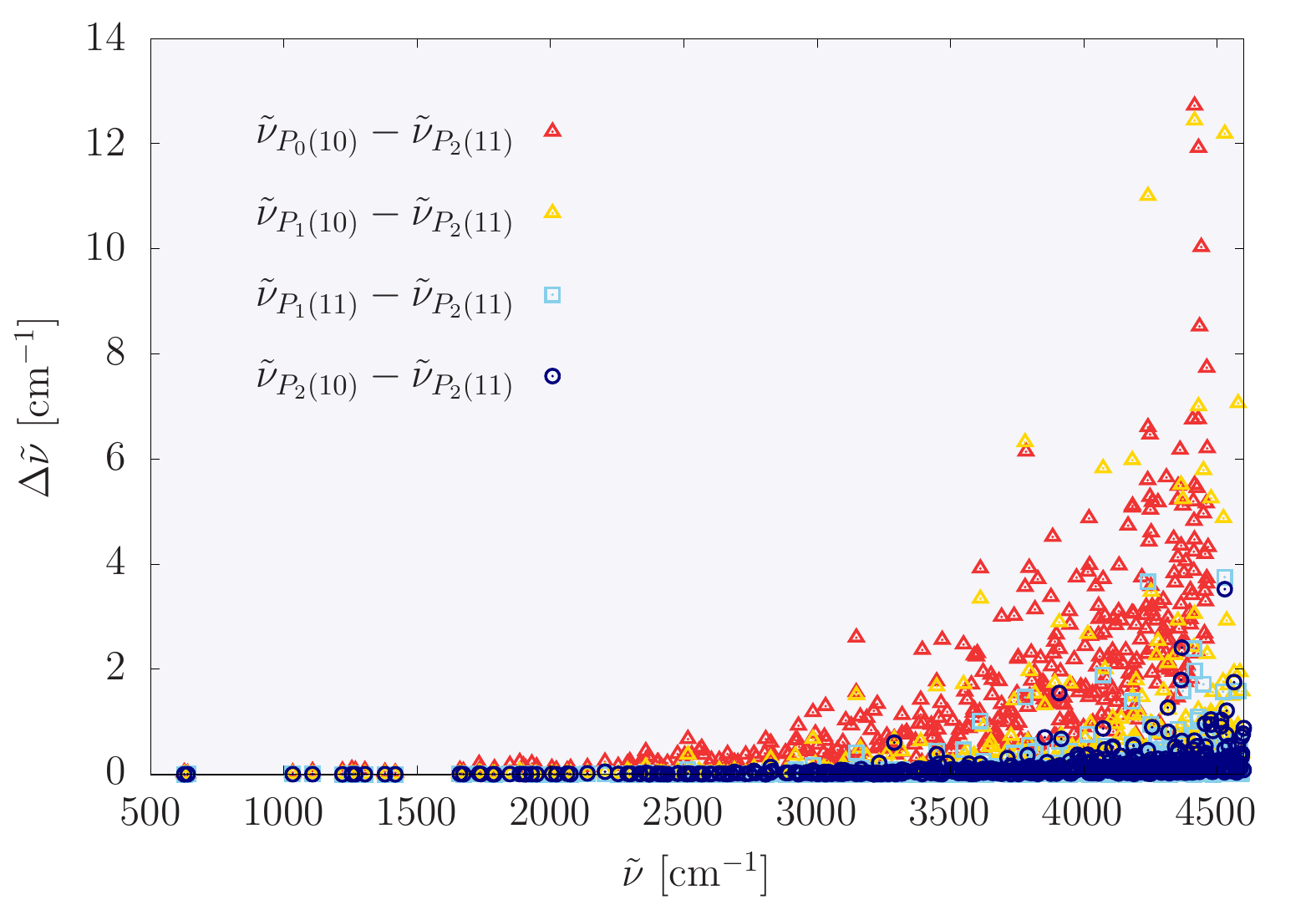}  
  \caption{%
    Convergence of the vibrational energies with respect to the largest computation, $P_2(b=11)$. The basis sets are labelled with $P_i(b),\ i=0,1,2$ according to details in Table~\ref{tab:prun}. 
    \label{fig:conver}
        }
\end{figure}

\subsection{Convergence of the vibrational states}
Convergence of all (\trans, \cis, \delocalized) vibrational states up to 4500~\cm\ beyond the zero-point energy have been tested in a series of computations (Fig.~\ref{fig:conver}). Table~\ref{tab:prun} collects parameters of three different (basis and grid) pruning conditions with different basis sizes. 

The figure shows that all vibrational energies of the $P^2(b=11)$ computation are converged within 4~\cm\ (up to 4500~\cm), and closer inspection of the data shows that most vibrational energies are much better converged (on the order of 0.01~\cm).

It can also be seen that the $P_2$ pruning condition, which generates a significantly larger basis than $P_1$ (Table~\ref{tab:prun}), is necessary to achieve good convergence beyond 3500~\cm. In particular, $P_1$ is insufficient, and the larger $P_2$ pruning is necessary, for good convergence of vibrational states with high $\nu_7$ (OCO bending) excitations.

\section{Computed vibrational spectra\label{sec:computed}}

\subsection{Vibrational infrared spectrum\label{sec:ir}}
In the present work, we focus on jet-cooled vibrational spectroscopy and
we approximate vibrational intensities using vibrational transition moments computed for the body-fixed properties (Sec.~\ref{sec:coordef}).
For computing vibrational spectra, the vibrational intensities can be approximated as\cite{SchReiMatCs09,WiDeCr80} 
\begin{align}
  &A^{\text{IR}}(\text{f} \leftarrow \text{i}) / (\text{km mol$^{-1}$}) \nonumber \\
  &\quad = 
  2.506562213
  [(\tilde\nu_\text{f} - \tilde\nu_\text{i}) / \text{cm$^{-1}$}] 
  \sum_{i = x,y,z} 
  \big[ |\bra{\psi_\text{f}} \mu_i \ket{\psi_\text{i}}|^2 / \text{Debye$^2$} \big] \; ,
\label{eq:vibint}
\end{align}
where $\tilde\nu_\text{i}$ and $\tilde\nu_\text{f}$, in \cm, correspond to the initial and final level energy, respectively. 
The $\bra{\psi_\text{f}} \mu_i \ket{\psi_\text{i}}$
transition dipole moment is calculated using the body-fixed dipole moment surface (DMS) and the $\psi_\text{f}$ final and $\psi_\text{i}$ initial vibrational wave functions.
The simulated jet-cooled spectrum (including transitions only from the vibrational ground state)
obtained with $\mu_i$ expressed in the Eckart frame of the \trans\ equilibrium structure
is shown in Fig.~\ref{fig:ir}. The spectrum obtained with the transition dipoles expressed in the body-fixed frame of Fig.~\ref{fig:coord} is provided in the \som\ (Fig.~S1), and it has similar features to Fig.~\ref{fig:ir}. 
Long ago, it was demonstrated by Le Sueur, Miller, Tennyson, and Sutcliffe that the frame dependence of the vibrational dipole transition moments may be significant,\cite{SuMiTeSu92} and it was emphasized that only the rovibrational transition moments are rigorously connected to the physical observables, independent of the mathematical details of the computation.

%
\begin{figure}
  \begin{center}
    \includegraphics[width=10cm]{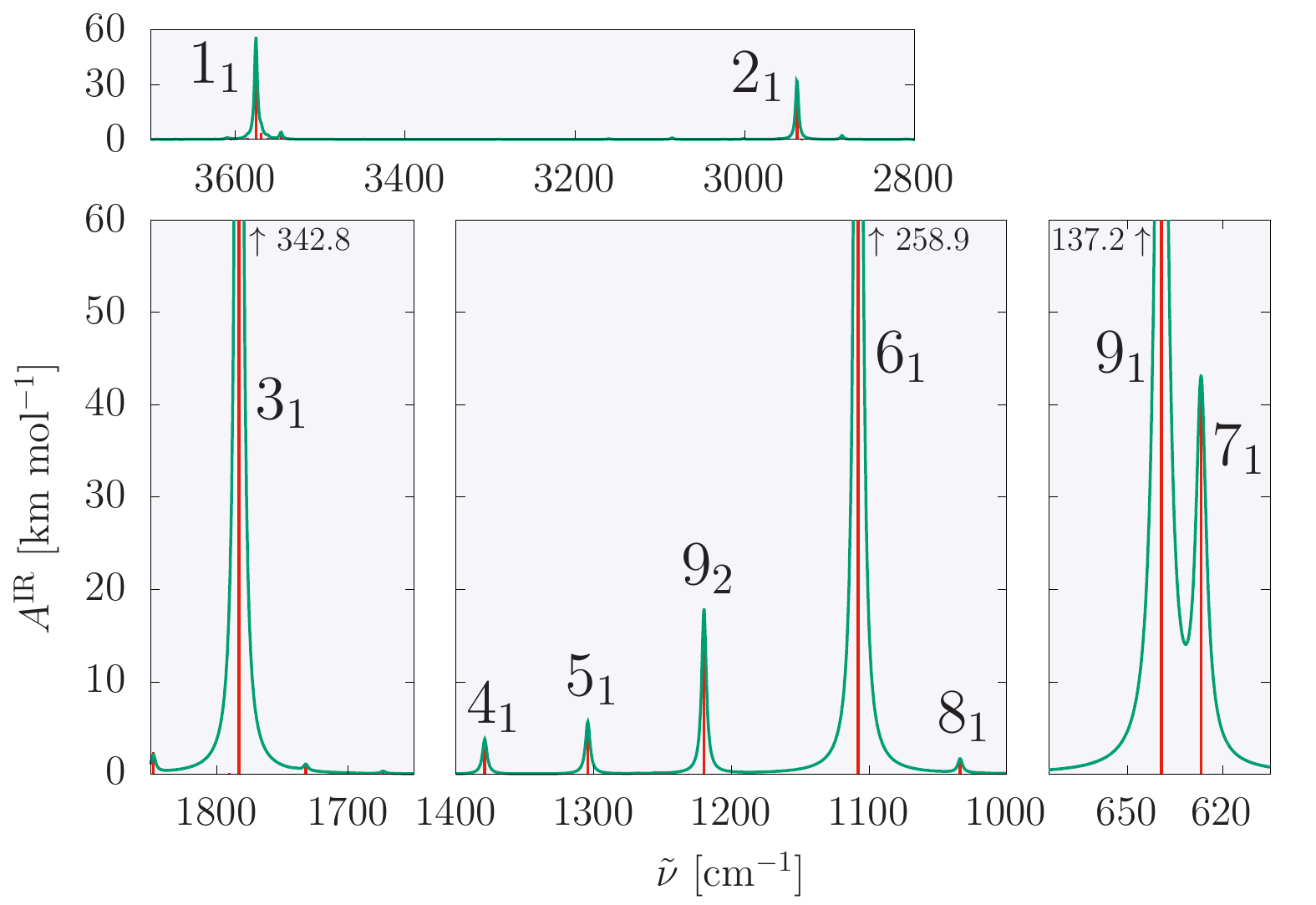}
  \end{center}    
  \caption{%
    Simulated vibrational infrared spectrum of HCOOH (transitions from the vibrational ground state).
    The vibrational intensities were computed with the Eckart frame corresponding to the \trans\ equilibrium structure.
    The stick spectrum (in red), computed according to Eq.~\eqref{eq:vibint}, has been convoluted (in green) using a Lorentz distribution with full-width-at-half-maximum (FWHM) of 5~\cm.
    \label{fig:ir}
    }
\end{figure}

\subsection{Vibrational Raman spectrum\label{sec:raman}}
To simulate the jet-cooled vibrational Raman spectrum, we compute the 
 parallel, perpendicular, and
total activity,\cite{Lo77,NeReKiHe02} 
\begin{align}
  A_{||}^{\text{R}}(\text{f} \leftarrow \text{i}) 
  &= 
  45 a_\text{fi}^2
  + 
  4 \gamma^2_\text{fi} \; ,
  \label{eq:ramanpara}
  \\
  A_{\perp}^{\text{R}}(\text{f} \leftarrow \text{i}) 
  &= 
  3 \gamma^2_\text{fi} \; ,
  \label{eq:ramanperp}
  \\
  A^{\text{R}}(\text{f} \leftarrow \text{i}) 
  &= 
  45 a_\text{fi}^2
  + 
  7 \gamma^2_\text{fi} \; ,
  \label{eq:vibraman}
\end{align}
respectively. 
It is necessary to note that this definition\cite{Lo77,NeReKiHe02} of the $A_{||}$ parallel and $A_\perp$ perpendicular
refers to the relative polarization of the incident and scattered photons.
For every band, the depolarization ratio is obtained as \cite{Lo77} 
\begin{align}
  \rho_{\text{fi}}
  =
  \frac{%
    A_{\perp}^{\text{R}}(\text{f} \leftarrow \text{i}) 
  }{%
    A_{||}^{\text{R}}(\text{f} \leftarrow \text{i}) 
  } \; .
\end{align}

\begin{figure}
  \begin{center}
    \includegraphics[width=10cm]{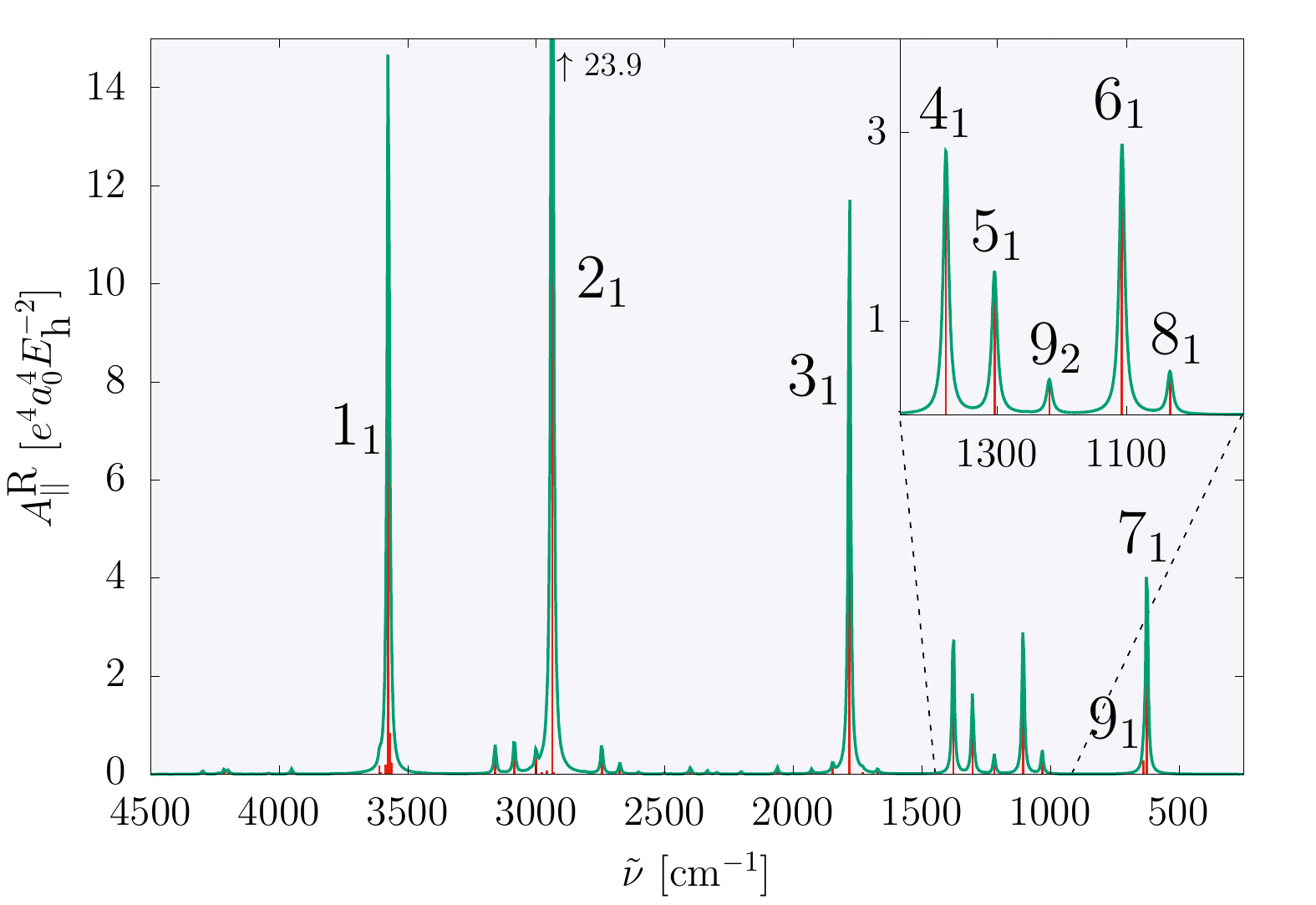}
  \end{center}    
  \caption{%
    Parallel Raman spectrum of HCOOH (transitions from the vibrational ground state). 
    The vibrational activities were computed with the Eckart frame corresponding to the \trans\ equilibrium structure.
    The stick spectrum (in red), Eq.~\eqref{eq:ramanpara}, is convoluted (in green) with a Lorentz distribution with FWHM of 10~\cm.
    \label{fig:raman_para}
    }
\end{figure}

The quantities $a_\text{fi}$ and $\gamma^2_\text{fi}$
are calculated according to Eqs.~\eqref{eq:pol_para} and \eqref{eq:pol_perp} using the vibrational transition moments for the polarizability components
\begin{align}
  (\alpha_{ij})_\text{fi}
  =
  \langle \Psi_\text{f} | \alpha_{ij} | \Psi_\text{i}\rangle \; ,
  \quad i,j=1(x),2(y),3(z).
\end{align}
Figures~\ref{fig:raman_para} and \ref{fig:raman_perp} show the simulated jet-cooled spectra {(assuming transitions only from the vibrational ground state)} obtained with $\alpha_{ij}$ expressed in the Eckart frame of the trans equilibrium structure. 
Figures~S2 and S3 of the \som\ show the same results but with $\alpha_{ij}$ expressed in the frame of Fig.~\ref{fig:coord}. It is interesting to note that both the parallel and perpendicular Raman bands (Figs.~\ref{fig:raman_para} and \ref{fig:raman_perp}) assigned to the $A''$ fundamental vibrations ($8_1$ and $9_1$, Table~\ref{tab:assign}) have a strong frame dependence (due to the large variation of $\gamma^2_\text{fi}$), while all other bands (of $A'$ symmetry) visible in these figures are only little affected by the frame change. Since these fundamental vibrations are localized in the \trans\ well and \cis-\trans\ delocalization has a negligible contribution, the \trans\ Eckart frame is expected to provide small rovibrational coupling, and hence, reasonably good vibrational activities, at least for the plotted energy range. In future work, we plan to compute the rovibrational states and the rigorous rovibrational transition moments, which are independent of the mathematical details (choice of the body-fixed frame) of the computation.

Table~\ref{tab:assign} collects the most intense peaks (the fundamentals of \trans-HCOOH and one overtone) in the infrared and Raman spectra. The full list of computed energies and transition moments is provided as \som.

\begin{figure}
  \begin{center}
    \includegraphics[width=10cm]{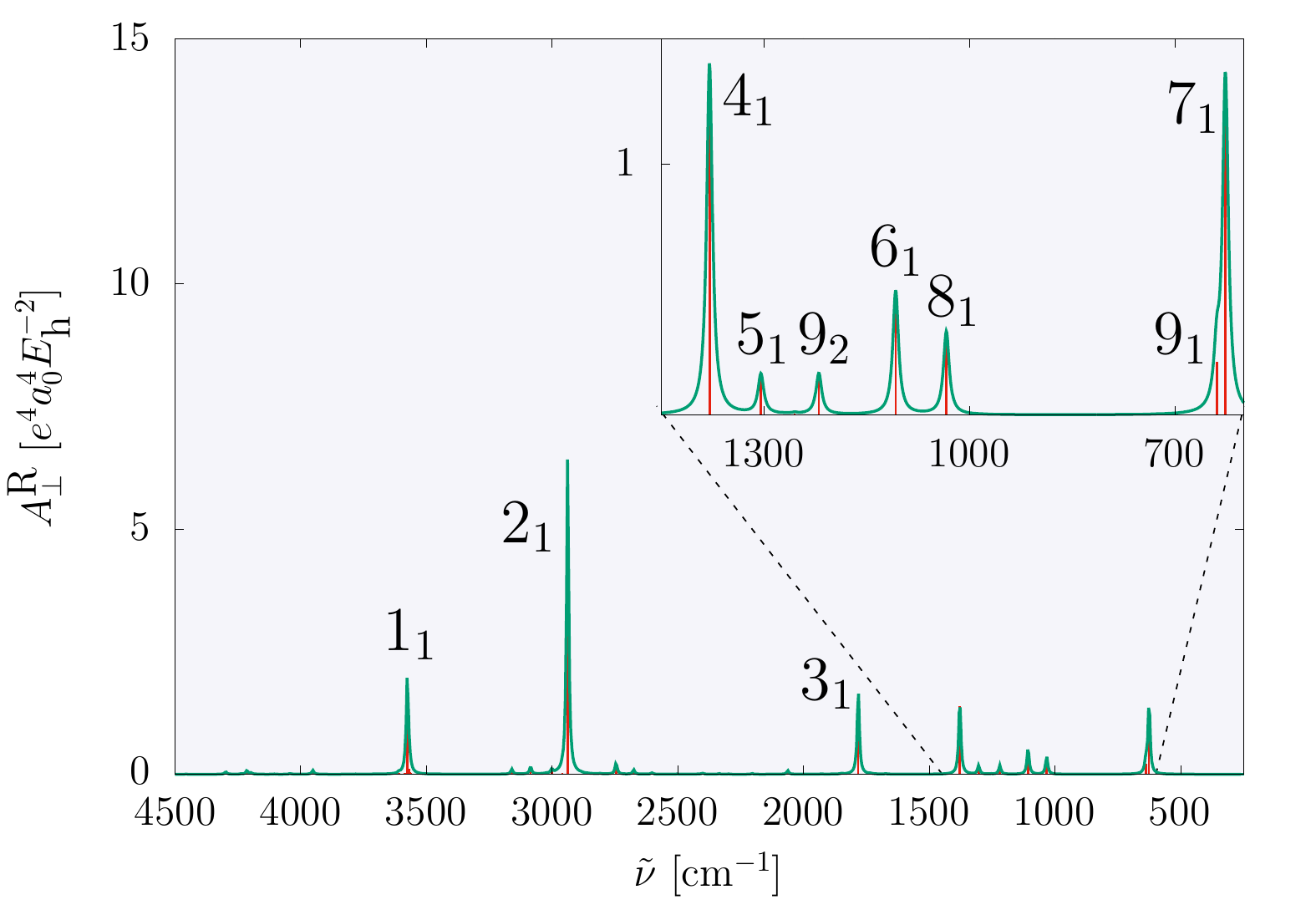}
  \end{center}    
  \caption{%
    Perpendicular Raman spectrum of HCOOH (including transitions from the vibrational ground state). 
    The vibrational activities were computed with the Eckart frame corresponding to the \trans\ equilibrium structure. 
    The stick spectrum (in red), Eq.~\eqref{eq:ramanperp}, is convoluted (in green) with a Lorentz distribution with FWHM of 10~\cm.
    \label{fig:raman_perp}
    }
\end{figure}

\begin{table}[h]
\caption{%
Vibrational energies referenced to the vibrational ground state, $\tilde\nu$ in \cm,  infrared intensities, $A^{\text{IR}}$ in km mol$^{-1}$, and total Raman activities, $A^{\text{R}}$ in $e^4 a_0^4 E_\text{h}^{-2}$, 
computed in this work with the GENIUSH-Smolyak program\cite{AvMa19,AvMa19b,DaAvMa22} and using the PES developed by Tew and Mizukami.\cite{TeMi16}
Gas-phase experimental data is also listed for comparison based on the collection of Ref.~\citenum{Ne22PhD}, the original reference is shown after each value.
\label{tab:assign}  
}
\centering
\scalebox{1.0}{
\begin{tabular}{@{}l@{\ \ } r@{\ \ \ }  r@{\ \ } r@{\ \ \ } c@{\ \ } r@{\ \ }r@{\ \ } c@{}}
\hline\\[-0.30cm]
\hline\\[-0.30cm]
 &  
 \multicolumn{3}{c}{Computed (this work)} && 
 \multicolumn{1}{c}{Obs. Raman} && 
 \multicolumn{1}{c}{Obs. IR} \\
\cline{2-4} \cline{6-6} \cline{8-8}\\[-0.30cm]
 Label & 
 \multicolumn{1}{c}{$\tilde\nu$} & 
 \multicolumn{1}{c}{$A^{\text{IR}}$} & 
 \multicolumn{1}{c}{$A^{\text{R}}$}
 && 
 \multicolumn{1}{c}{$\tilde\nu$} &&
 \multicolumn{1}{c}{$\tilde\nu$} \\
\\[-0.30cm]
\hline\\[-0.30cm]
$7_1$ $(A')$  &     626.8    &  41.7  &  5.3  &&   626\cite{NeSi21}   && 626.17\cite{ref210} \\
$9_1$ $(A'')$ &     639.4    & 137.2  &  0.5  &&         &&    640.73\cite{ref210}  \\
$8_1$ $(A'')$ &     1033.8   &   1.5  &  0.8  &&  1033\cite{NeSi21}   && 1033.47\cite{ref204}  \\
$6_1$ $(A')$ &     1107.8   & 258.9  &  3.3  &&  1104\cite{NeSi21}   &&   1104.85\cite{ref204}  \\
$9_2$ $(A')$ &     1219.9   &  17.95  &  0.5  &&  1220\cite{NeSi21}   &&   1220.83\cite{ref61}  \\
$5_1$ $(A')$ &     1304.4   &   5.7  &  1.6  &&  1306\cite{NeSi21}   &&   1306.14\cite{ref47} \\
$4_1$ $(A')$ &   1379.2     &   3.8  &  4.1  &&  1379\cite{NeSi21}   &&   1379.05\cite{ref218}  \\
$3_1$ $(A')$ &   1782.8     & 342.8  & 12.6  &&  1776\cite{NeSi21}   &&   1776.83\cite{ref48}  \\
$2_1$ $(A')$ &   2938.3     &  32.0  & 29.7  &&  2942\cite{NeSi21}   &&   2942.06\cite{ref46}  \\
$1_1$ $(A')$ &   3575.8     &  53.7  & 15.0  &&  3570\cite{NeSi21}   &&   3570.5\cite{ref46}   \\
\hline\\[-0.30cm]
\hline\\[-0.30cm]
\end{tabular}
}
\end{table}

\section{Summary and conclusions\label{sec:concl}}
\noindent
Variational vibrational states and infrared and Raman transition moments have been computed for the formic acid molecule.
Basis- and grid-pruning conditions have been developed in order to converge all (including \trans, \cis, and \delocalized) vibrational states with the GENIUSH-Smolyak computer program\cite{AvMa19,AvMa19b,DaAvMa22} and using the \emph{ab initio} PES of Ref.~\citenum{TeMi16}. 
All vibrational energies are converged within 4~\cm\ up to 4500~\cm\ beyond the vibrational ground state.
Most of the vibrational energies are much better converged (within 0.01~\cm), the $\nu_7$ OCO bending mode appears to be the most anharmonic mode among the small-amplitude degrees of freedom and requires large basis sizes for good convergence.

For direct comparison with experimentally recorded infrared and Raman spectra, full-dimensional electric dipole and dipole polarizability surfaces have been developed (for the coordinate range relevant for the quantum dynamics) using \emph{ab initio} data points obtained at the CCSD/aug-cc-pVTZ level of theory with the Dalton program package.\cite{Dalton}

The variational vibrational states and the developed property surfaces were used to compute (body-fixed) dipole and polarizability transition moments, and these transition moments were used to simulate jet-cooled vibrational infrared and Raman spectra. Further analysis of the results and comparison with the PES of Ref.~\citenum{RiCa18} for the parent as well as the isotopologue species is left for further work. 
Similarly, rovibrational energies and transition moments will be reported in the future.

\section*{Conflicts of interest}
There are no conflicts to declare.

\section*{Acknowledgements}
We thank Arman Nejad for discussions about the experimental vibrational Raman spectrum, which led us to discover the strong frame dependence of the $8_1$ and $9_1$ $A''$ fundamental bands. 
We also thank the financial support of 
the Hungarian National Research, Development, and Innovation Office (FK~142869).

\clearpage

%

\clearpage

\setcounter{section}{0}
\renewcommand{\thesection}{S\arabic{section}}
\setcounter{subsection}{0}
\renewcommand{\thesubsection}{S\arabic{section}.\arabic{subsection}}

\setcounter{equation}{0}
\renewcommand{\theequation}{S\arabic{equation}}

\setcounter{table}{0}
\renewcommand{\thetable}{S\arabic{table}}

\setcounter{figure}{0}
\renewcommand{\thefigure}{S\arabic{figure}}

~\\[0.cm]
\begin{center}
\begin{minipage}{0.8\linewidth}
\centering
\textbf{Supplementary Information} \\[0.25cm]

\textbf{%
Vibrational infrared and Raman spectrum of HCOOH from variational computations}
\end{minipage}
~\\[0.5cm]
\begin{minipage}{0.9\linewidth}
\centering

Gustavo Avila,$^1$ Alberto Mart\'in Santa Dar\'ia,$^{1,2}$ and Edit M\'atyus$^{1,\ast}$ \\[0.15cm]

$^1$~\emph{ELTE, Eötvös Loránd University, Institute of Chemistry, 
Pázmány Péter sétány 1/A, Budapest, H-1117, Hungary} \\[0.15cm]
$^2$~\emph{Departamento de Química Física, University of Salamanca,
37008 Salamanca, Spain} \\[0.15cm]
$^\ast$ edit.matyus@ttk.elte.hu \\
\end{minipage}
~\\[0.15cm]
(Dated: April 14, 2023)
\end{center}


\setcounter{section}{0}
\renewcommand{\thesection}{S\arabic{section}}
\setcounter{subsection}{0}
\renewcommand{\thesubsection}{S\arabic{section}.\arabic{subsection}}

\setcounter{equation}{0}
\renewcommand{\theequation}{S\arabic{equation}}

\setcounter{table}{0}
\renewcommand{\thetable}{S\arabic{table}}

\setcounter{figure}{0}
\renewcommand{\thefigure}{S\arabic{figure}}


~\\[2cm]

\begin{figure}[h]
  \begin{center}
    \includegraphics[width=10cm]{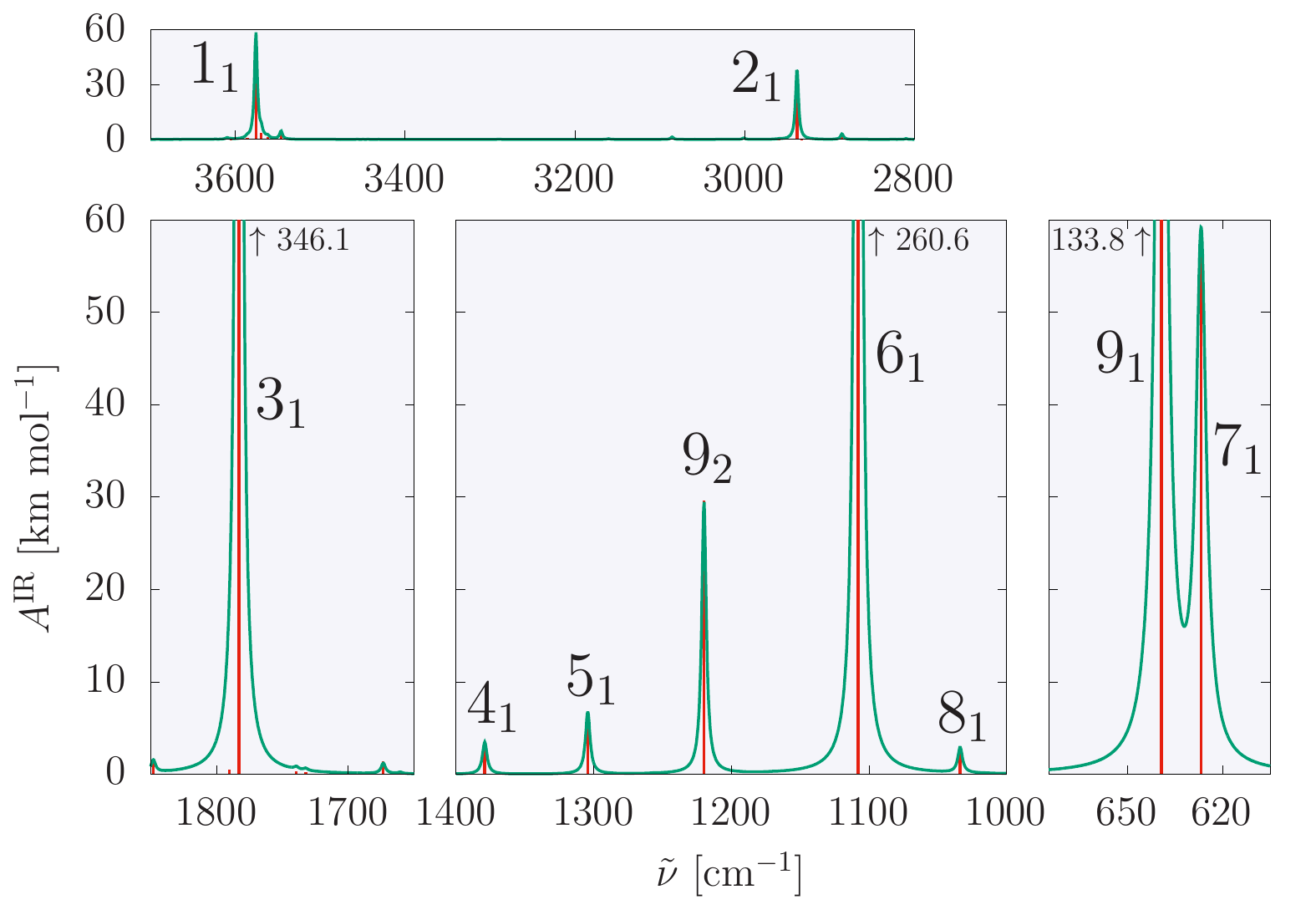}
  \end{center}    
  \caption{%
    Simulated vibrational infrared spectrum of HCOOH (transitions from the vibrational ground state) corresponding to the body-fixed frame of Fig.~1. 
    The stick spectrum (in red) has been convoluted (in green) using a Lorentz distribution with full-width-at-half-maximum (FWHM) of 5~\cm.
    It is worth comparing this figure with Fig.~8 of the manuscript, which corresponds to the Eckart frame of the \trans\ equilibrium structure, small differences in the vibrational intensity values can be observed.
    \label{fig:ir}
    }
\end{figure}

\begin{figure}[h]
  \begin{center}
    \includegraphics[width=10cm]{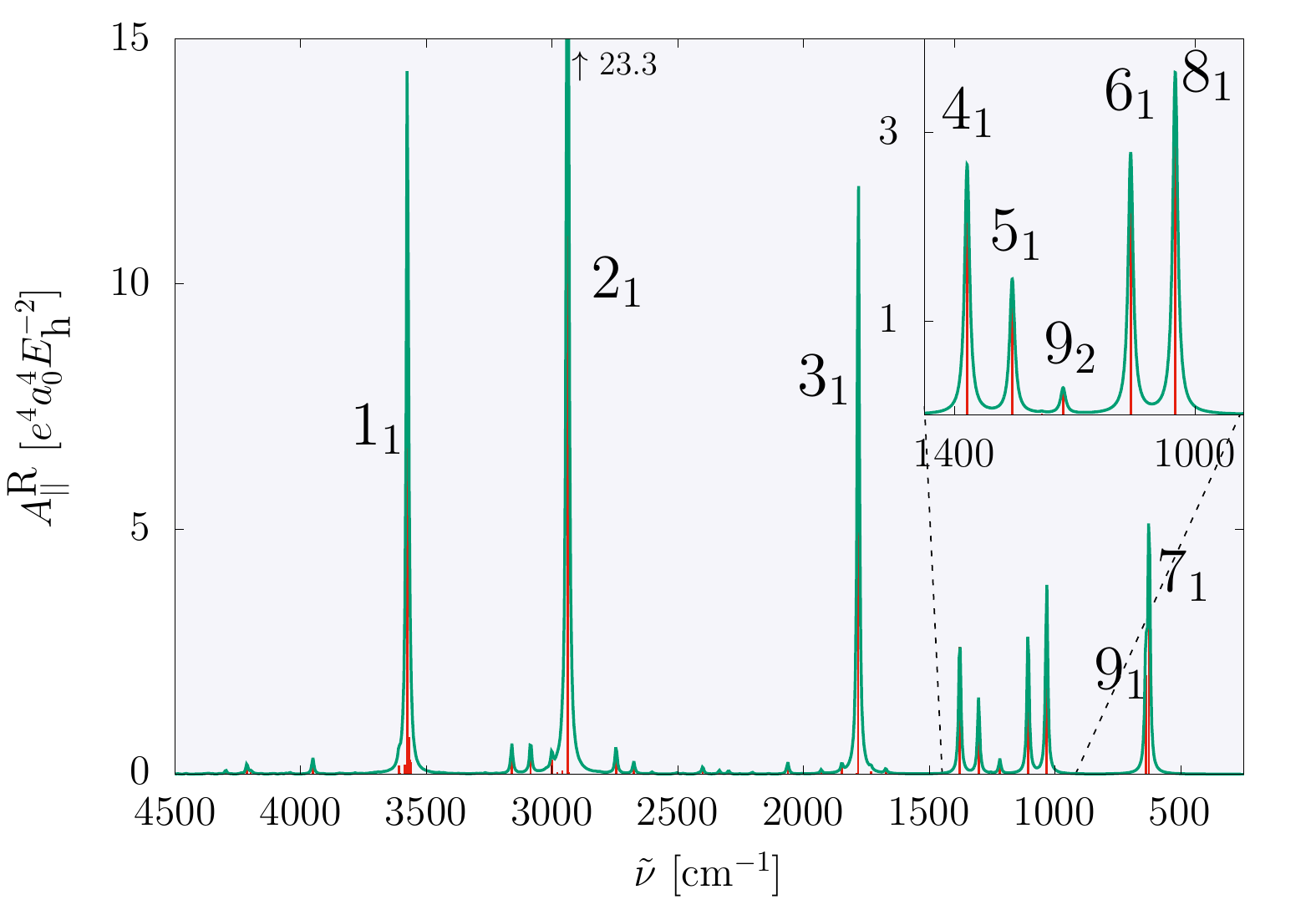}
  \end{center}    
  \caption{%
    Parallel Raman spectrum of HCOOH (including transitions from the vibrational ground state) corresponding to the body-fixed frame of Fig.~1 of the manuscript. 
    The stick spectrum (in red) is convoluted (in green) with a Lorentz distribution with FWHM of 10~\cm.
    It is worth comparing this figure with Fig.~9 of the manuscript, which corresponds to the Eckart frame of the \trans\ equilibrium structure. Major (ca. an order-of-magnitude) difference in the $8_1\ (A'')$  and $9_1\ (A'')$ activities can be observed, which is discussed in Sec. 5.
    \label{fig:raman_perp}
    }
\end{figure}

\begin{figure}[h]
  \begin{center}
    \includegraphics[width=10cm]{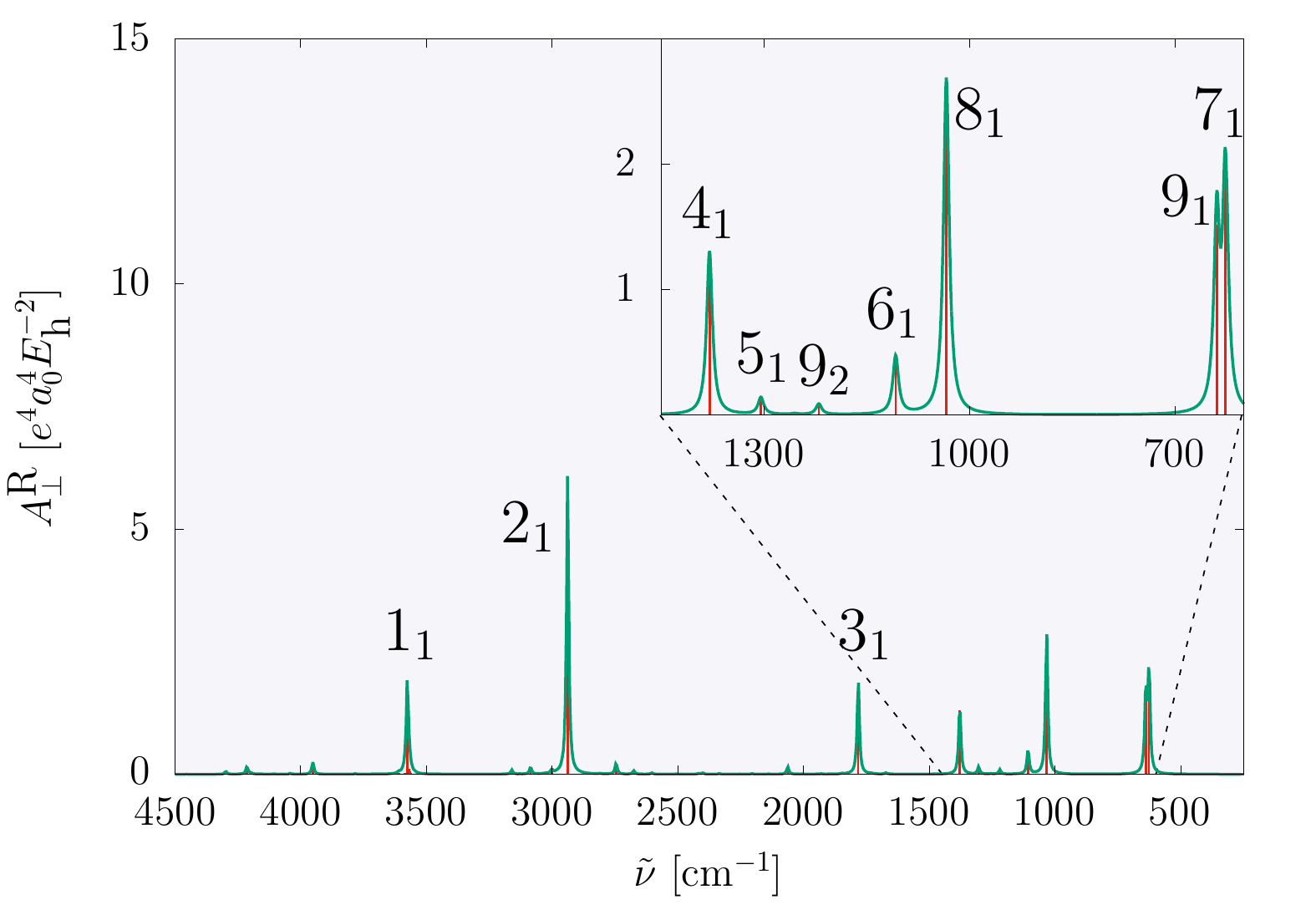}
  \end{center}    
  \caption{%
    Perpendicular Raman spectrum of HCOOH (including transitions from the vibrational ground state) corresponding to the body-fixed frame of Fig.~1 of the manuscript. 
    The stick spectrum (in red) is convoluted (in green) with a Lorentz distribution with FWHM of 10~\cm.
   It is worth comparing this figure with Fig.~10 of the manuscript, which corresponds to the Eckart frame of the \trans\ equilibrium structure.  Major (ca. an order-of-magnitude) difference in the $8_1\ (A'')$  and $9_1\ (A'')$ activities can be observed, which is discussed in Sec. 5.
    \label{fig:raman_perp}
    }
\end{figure}

\end{document}